# Transformative Science from the Lunar Farside: Observations of the Dark Ages and Exoplanetary Systems at Low Radio Frequencies

**Jack O. Burns***

*Center for Astrophysics and Space Astronomy, University of Colorado Boulder, Boulder, CO 80309 USA*

**Keywords:** lunar radio telescopes, 21-cm cosmology, exoplanet magnetospheres, low radio frequency observations

## Summary

The farside of the Moon is a pristine, quiet platform to conduct low radio frequency observations of the early Universe's Dark Ages, as well as space weather and magnetospheres associated with habitable exoplanets. In this paper, the astrophysics associated with NASA-funded concept studies will be described including a lunar-orbiting spacecraft, DAPPER, that will measure the 21-cm global spectrum at redshifts ≈40-80, and an array of low frequency dipoles on the lunar farside surface, FARSIDE, that would detect exoplanet magnetic fields. DAPPER observations (17-38 MHz), using a single cross-dipole antenna, will determine the amplitude of the 21-cm spectrum to the level required to distinguish the standard ΛCDM cosmological model from those produced by exotic physics such as nongravitational dark matter interactions. FARSIDE has a notional architecture consisting of 128 dipole antennas deployed across a 10 km area by a rover. FARSIDE would image the entire sky each minute in 1400 channels over 0.1-40 MHz. This would enable monitoring of the nearest stellar systems for the radio signatures of coronal mass ejections and energetic particle events, and would also detect the magnetospheres of the nearest candidate habitable exoplanets. In addition, FARSIDE would provide a pathfinder for power spectrum measurements of the Dark Ages.

## 1. Introduction

The lunar farside is the only truly radio-quiet zone in the inner solar system. The Earth's surface at frequencies ≲50 MHz is contaminated by anthropogenic radio frequency interference (RFI) from heavily-used, powerful civil and military transmitters. In addition, the Earth's ionosphere corrupts low frequency total power measurements via absorption (with the ionosphere becoming opaque below ~10 MHz), emission, and refraction driven in large part by solar emissions and the solar wind (see [1] and references therein). Equally important are the interactions of the radio antenna beam with the ground, ground screens, and horizon features which complicate the beam chromaticity and the extraction of the beam-averaged foreground from weak signals [2].

Observations in Earth orbit or in cis-lunar space eliminate some but not all of the issues required for a truly radio-quiet environment. As shown in Figure 1, even at the distance of the Moon, RFI is severe even when not shielded from behind the Moon. The equivalent brightness temperature from human-generated RFI at the distance of the Moon at ~17 MHz is ~750,000 K (compared to the ~100 mK expected brightness of the redshifted 21-cm background; [3]). On the other hand, NASA's RAE-2 confirmed that above the lunar farside, the Galaxy's low frequency emission is easily detected [4] and serves as the foreground through which cosmological measurements must be made.

*Author for correspondence: J. Burns (jack.burns@colorado.edu).

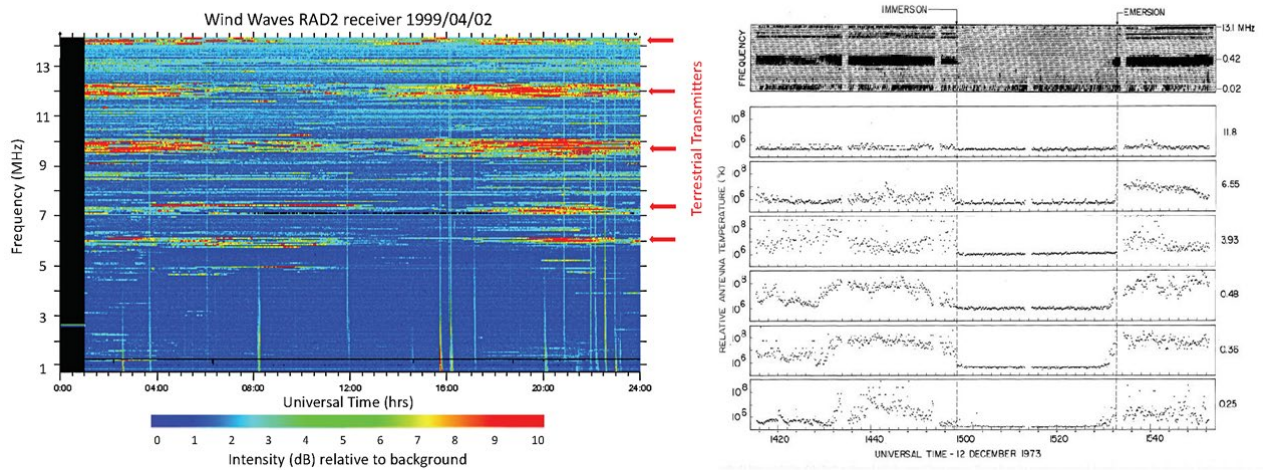

**Figure 1**. *Left:* Low radio frequency observations collected by the Wind/Waves spacecraft when it was in the vicinity of the Moon illustrate the high level of RFI from Earth (figure courtesy of R. MacDowall, GSFC). *Right:* Measurements from NASA's RAE-2 satellite demonstrate the radio-quiet zone above the lunar farside [4].

The advantages of the Moon's farside for low radio frequency observations were recognized even before the launch of the Apollo missions [5]. Over the subsequent ≈50 years, a number of concepts for low frequency astronomy at or on the Moon have been proposed (see e.g., [6,7]). These include single antennas and arrays in lunar orbit as well as radio interferometers on the lunar farside [8]. Early science proposed from the Moon included nonthermal radiation from plasma instabilities within the solar system, observations of ionized hydrogen in the interstellar medium (ISM) via absorption at ≲5 MHz, and measurements of the low frequency spectra of extragalactic radio sources [6]. However, the early science cases lacked a "killer App", that is, a compelling astrophysics problem that uniquely requires low frequency observations from the Moon.

Over this past decade, two exciting applications have emerged in the areas of redshifted 21-cm cosmology as well as exoplanet magnetospheres and habitability. In this paper, I will describe lunar mission concepts to observe hydrogen in the Dark Ages of the early Universe (redshifts z∼40-80) to constrain physics beyond the standard ΛCDM cosmology using lunar orbiting and lunar surface dipole antennas. In addition, I will present a new concept for a practical radio interferometric array on the farside to observe energetic particle events and the magnetospheres associated with exoplanetary systems at frequencies down to ~100 kHz. This paper will focus on the science goals for each telescope with only summaries presented for the instrument designs.

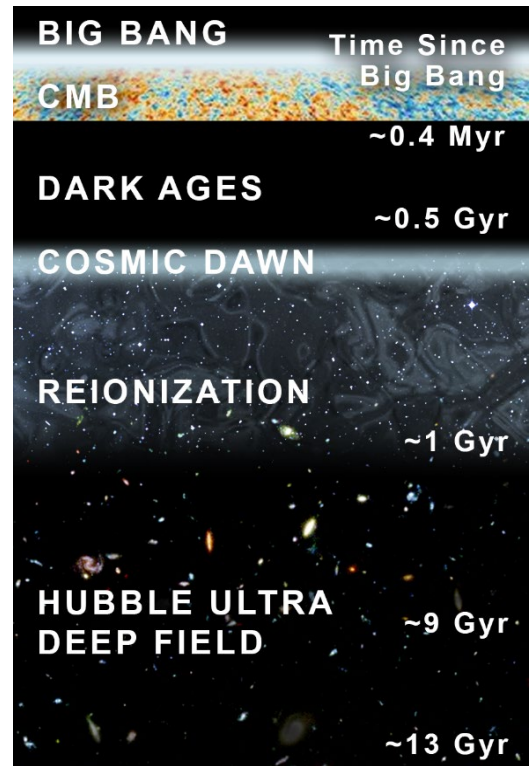

**Figure 2.** The pre-stellar (Dark Ages), first stars (Cosmic Dawn), and Reionization epochs of the Universe can be uniquely probed using the redshifted 21-cm signal. This history is accessible via the neutral hydrogen spin-flip background [9]. Credit: Robin Clarke, Caltech/ JPL.

## 2. The Dark Ages global 21-cm spectrum

There is an enormous gap in our understanding of the early Universe between the Epoch of Recombination ($z≈1100$, ~0.4 Myr after the Big Bang), probed via the Cosmic Microwave Background (CMB), and the later stages of the Epoch of Reionization ($z∼6$-10, ~1 Gyr), sampled by HST. This void of observations begins at the Dark Ages, with the intergalactic medium (IGM) composed of newly formed neutral hydrogen, extends through Cosmic Dawn, when the first stars, galaxies, and accreting black holes form at $z∼30$-40, and ends in the Universe's last phase transition with an ionized IGM (see Figure 2). Models of ΛCDM cosmology and primordial star formation are untested during this critical time representing the birth of structural complexity in the Universe. There is one, unique way to probe these early epochs, however, using the redshifted 21-cm background observed from the farside of the Moon.

### (a) The hydrogen cosmology signal

The spin-flip, hyperfine transition of neutral hydrogen produces emitted radio emission at 21-cm (1420 MHz) which is redshifted by a factor of $(1 + z)$ due to the expansion of the Universe (i.e., $\nu_{observed}$=17 MHz at $z$=80). Since neutral hydrogen composed the majority of baryonic matter during the Dark Ages and Cosmic Dawn, its global 21-cm spectrum can be detected even by a single dipole antenna above or on the lunar farside. Figure 3 shows models and a claimed detection of the Cosmic Dawn absorption trough in the global 21-cm spectrum. The Dark Ages trough is purely cosmological and thus relatively simple since there are no stars or galaxies during this epoch. The standard cosmology model (black dashed curve) makes a precise prediction of the frequency and $\delta T_b$ (brightness temperature measured relative to the radio background) of the minimum of the trough since only adiabatic expansion cools the gas relative to the CMB. Any deviation from this prediction is evidence of new physics.

The higher frequency Cosmic Dawn trough is produced by the onset of the first stars. These stars radiate prodigious amounts of Ly$\alpha$ photons that couple with the hyperfine transition of neutral hydrogen and produce an absorption feature via the Wouthuysen-Field effect [9]. X-ray heating of the IGM from the first accreting black holes, probably remnants of the first stars, produces the extremum at the bottom of the trough. The frequencies of the first inflection point and extremum of the trough are measures of the redshifts when stars and black holes, respectively, first turn on, and the shape constrains the mass of the first stars and black holes [1].

Bowman et al. [11] recently claimed a possible detection of the Cosmic Dawn trough using the EDGES instrument in western Australia. As shown in Figure 3, the EDGES results differ quite significantly from the expectations of the standard model. Although the frequency of the bottom of the trough can be understood within the uncertainties of the standard model of star formation, the depth is much greater than expected. It is well below what was thought to be a hard limit set by adiabatic expansion. This could be explained by an increase in the radio background above the CMB produced, for example, by synchrotron emission from early star-forming galaxies or AGNs [12], or dark matter annihilation [13]. Alternatively, the enhanced trough could be caused by added cooling of baryons via Rutherford-like scattering by partially-charged dark matter [14,15]. In either case, new, likely exotic physics, is required if this EDGES trough is confirmed.

Figure 3 illustrates parametric added hydrogen cooling models with amplitudes consistent with the EDGES results. The Cosmic Dawn trough originates from a combination of complex physics and astrophysics. On the other hand, the Dark Ages trough is much simpler since this epoch lacks astrophysics. The cooling models nicely separate at these lower frequencies and constrain deviations from standard ΛCDM cosmology that are caused by new physics such as unexpected interaction with dark matter.

A large number of papers have now been published on possible theoretical explanations of the enhanced absorption claimed by EDGES. But, there are good reasons to be skeptical of these EDGES results based upon possible instrumental systematics [16] or beam chromaticity or other unmodelled effects [17 and references therein]. Even so, observations of the redshifted 21-cm line in the Dark Ages have the potential for discovery of additional new physics in an unobserved epoch. This includes the decay and/or annihilation of dark matter [18], primordial black holes [19], cosmic strings [20], and early Dark Energy [21].

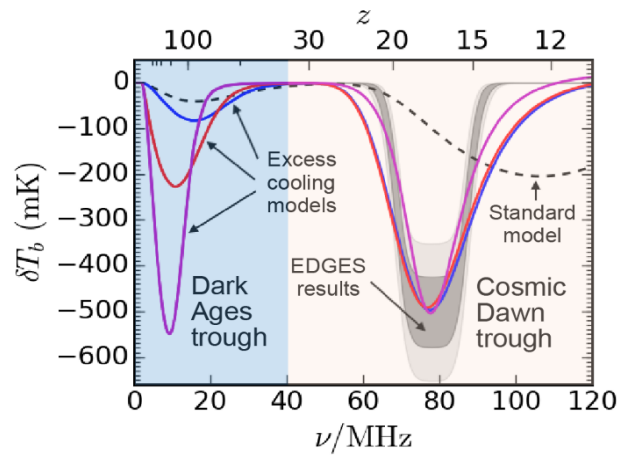

**Figure 3.** The global 21-cm spectrum provides a key test of standard ΛCDM cosmology and possible exotic physics produced by interactions with dark matter, as well as constraints on the properties of primordial stars. The black dashed curve is a prediction using standard cosmology with adiabatic hydrogen gas cooling and star formation similar to that in the Milky Way. The color curves are parametric models with added gas cooling [10]. The grey curves are 1- and 2-$\sigma$ uncertainties from EDGES [11] ground-based observations.

## (b) **DAPPER**: The Dark Ages Polarimeter Pathfinder

The DAPPER mission is designed to precisely measure the two absorption troughs in the redshifted 21-cm spectrum. It will resolve current ambiguities and cleanly constrain the origin and characteristics of added baryonic cooling and/or an enhanced radio background. The scientific objectives of DAPPER include:

1. Determine the level of (dis)agreement with the standard cosmological model possibly caused by dark matter in the Dark Ages.
2. Determine (a) the level of excess cooling above the adiabatic limit for Cosmic Dawn and (b) when the first stars and black holes formed.

To accomplish these goals, DAPPER will measure the hydrogen cosmology spectrum from ≈17 to 107 MHz (83≥ $z$ ≥12) with an rms thermal noise level of ≈20 mK.

These instrument functional requirements are met by a relatively simple cross-dipole antenna + spectropolarimeter placed in a low lunar orbit (50×100 km), with data taken when the spacecraft is in the radio shadow above the farside of the Moon [3]. The instrument components have heritage from previous space missions and ground-based prototypes.

The DAPPER antenna encompasses four deployable wire boom monopoles, arranged in two, orthogonal co-linear pairs as shown in Figure 4. At the deployment length of 7.57-m tip to tip, the antenna system corresponds to a half-wave dipole at 19 MHz. The antenna beam has a large solid angle but relatively little chromaticity (Figure 4).

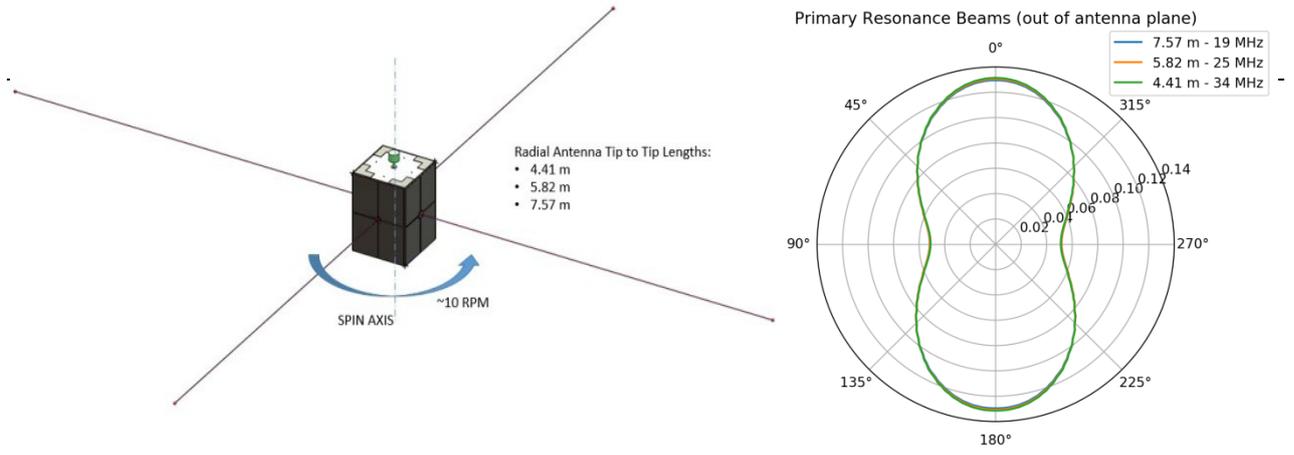

**Figure 4.** *Left:* DAPPER wire boom antennas shown fully deployed. The 0.8-m length spacecraft will spin to keep the antennas rigid and to perform dynamic polarimetry on the foreground [23]. *Right:* The cross-dipole antenna beam samples large areas of the sky with relatively little chromaticity across the low frequency band.

The radio receiver design is based in part on the FIELDS instrument currently flying on the *Parker Solar Probe* [22] and the instrument testbed for the *Cosmic Twilight Polarimeter* [23]. A calibration frequency tone generator has been added to the design to track the gain and reflection coefficient to 5 ppm/sec as required to achieve a high dynamic range measurement of the 21-cm signal in the presence of bright foregrounds.

The greatest challenge for DAPPER will be to separate the expected 21-cm spectrum from the beam-weighted foreground which is $\sim 10^{4\text{-}6}$ times more intense [1]. To accomplish this, we take advantage of the differences in spectral shapes, spatial structure, and polarization between the signal and the foreground. The foreground is produced by synchrotron radiation with a simple power-law spectrum whereas the 21-cm spectrum has the absorption troughs shown in Figure 3. The 21-cm background is isotropic when viewed with the large DAPPER antenna beam but the foreground has structure of a disk and bulge due to the Galaxy. This Galaxy structure produces an induced polarization when observed with a dipole antenna whereas the 21-cm background has only a Stokes I signal.

Telescopic observations of the foreground produce a weighting of the intrinsically smooth power-law spectrum by a frequency-dependent beam. As a result, chromaticity in the beam induces significant artifacts in the observed spectrum of the foreground which could be misinterpreted as structure in the cosmological 21-cm spectrum. Analyses involving multiple linear foreground models do not properly address the chromaticity issue and are unlikely to produce unique 21-cm spectra [17]. To address this problem, the chromatic beam, the foreground, and 21-cm spectrum must be modeled together.

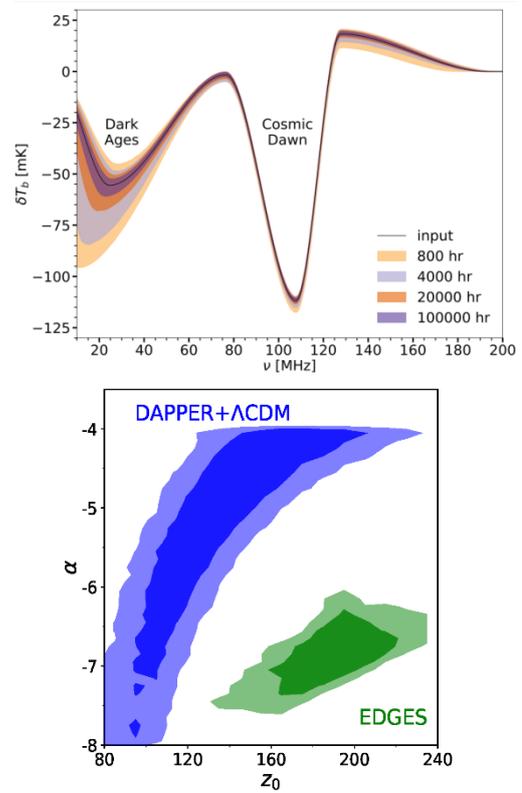

**Figure 5.** *Top:* Results for simulated DAPPER observations using our SVD/MCMC pipeline including statistical plus systematic uncertainties for four evenly increasing integration times from 800-100,000 hours [25]. *Bottom:* DAPPER will distinguish at >5σ between a standard ΛCDM model (blue) and an added cooling model as suggested by EDGES (green). $\alpha$ is the cooling rate and $z_0$ is the redshift where cooling begins.

Such an approach has recently been devised and extensively tested [24,25,26]. The DAPPER data analysis pipeline uses pattern recognition and Bayesian inference to account for experimental uncertainties including statistical thermal noise, produced by the foreground that dominates the system temperature at these frequencies, and systematic effects that change how the signal appears in the observed spectrum. The combination defines the likelihood function from which we can obtain the probability that a given set of model parameters describes the underlying physics of the observation. Training sets are used to constrain the possible variations in instrument, beam-weighted foreground, and 21-cm signal characteristics. For the pattern recognition segment of the pipeline, we perform Singular Value Decomposition (SVD) to encapsulate the eigenmodes of variation for each component forming the observational data set [24]. Unlike the traditional approach of fitting relatively generic polynomials and using a single averaged spectrum, our approach utilizes optimal basis vectors derived from detailed modelling training sets and multiple correlated spectra to avoid systematic biases to the level required to extract the predicted global 21-cm signal [26]. Given a nonlinear model of the signal, we then use Bayes' theorem by numerically sampling the likelihood function to calculate the posterior probability distribution of its parameters via a Markov Chain Monte Carlo (MCMC) algorithm, while analytically marginalizing over the weights of the SVD beam-weighted foreground eigenmodes and suppressing those unimportant based on priors gleaned from the training set [25]. An example of the results from our pipeline applied to a simulated DAPPER observation is shown in the top panel of Figure 5, which illustrates that DAPPER will measure the characteristics of both the Dark Ages and Cosmic Dawn spectral troughs with precision. This will result in a clean separation between the standard ΛCDM cosmological model and new physics models with added cooling as represented by the EDGES results, possibly produced by interactions with dark matter, at a level >5σ (see bottom panel of Figure 5).

# 3. The magnetospheres and space environments of habitable exoplanets

The discovery of extraterrestrial life would be one of humankind's greatest intellectual achievements. TESS is in the process of identifying Earth-like planets within the habitable zone for nearby exoplanetary systems. Further characterization, including spectral observations of atmospheric constituents, will be pursued by *JWST*, *WFIRST*, and possibly by *HabEx* and/or *LUVOIR*.

It is becoming clear that we must also consider the space weather environs of the host star and if the exoplanet has a large-scale magnetic field to protect it from energetic space plasma events originating from the star. Measurements by NASA's MAVEN mission have shown that the solar wind stripped much of Mars' atmosphere after the planet's core cooled and shut down the planet-wide magnetic field. Compared to Earth and its liquid water oceans, Mars is desolate without any sign of surface life. Thus, to determine if an exoplanet is truly habitable, we may also need to know the frequency and magnitude of space weather events within the planetary system and the strength of the magnetic field of the planet.

About 75% of stars out to a distance of 8 pc are estimated to be M dwarfs [27]. The most promising candidates for habitable planets likely orbit around these small cool stars. *Kepler* has shown that most M dwarfs harbor small planets, with 2.5±0.2 planets per star, with <4 Earth radii, and with orbital periods <200 days [28]. Thus, M dwarfs would appear to be good hunting grounds for potentially habitable planets. As shown schematically in Figure 6, the problem is that these stars are also known to be much more active, flaring often and with higher energy than what is observed from solar flares [29]. Since the habitable zone around M dwarfs is ≤0.2 AU [30], the effects of such flares and coronal mass ejections (CMEs) on planets will detrimentally impact the retention of their atmospheres and, thus, surface water. Magnetospheres can help shield planetary atmospheres from the catastrophic loss due to energetic particle events. So, to fully assess if an exoplanet has the conditions necessary for life, we need to measure the stellar weather and the planetary magnetic field in these systems.

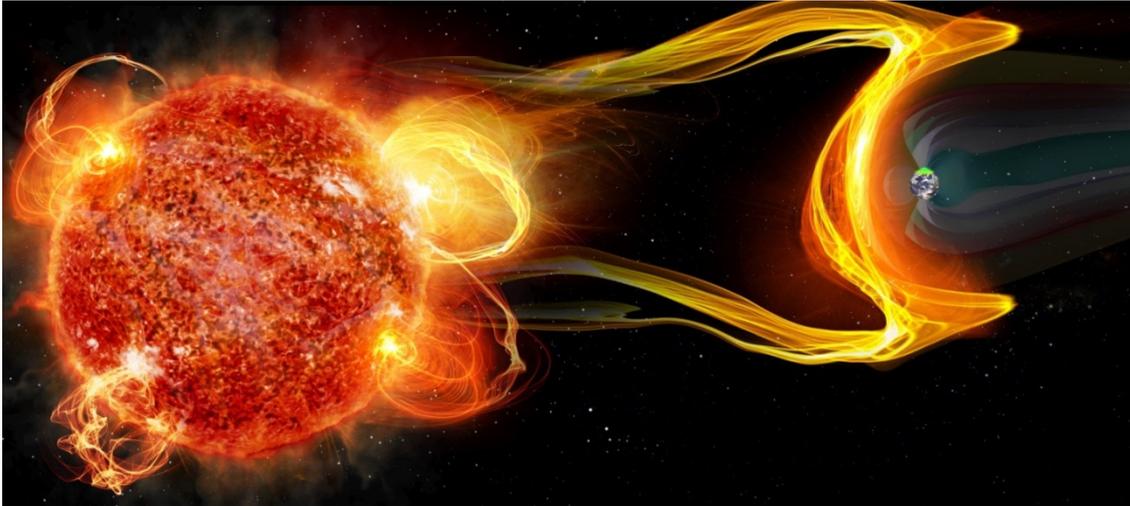

**Figure 6.** An artist's impression of an Earth-like planet experiencing a CME from an active M dwarf star. Detecting CMEs, energetic particle events, and magnetospheres of candidate habitable planets is a principal goal for the lunar FARSIDE low radio frequency array. Credit: Chuck Carter / Caltech / KISS.

### (a) Probing space weather in extrasolar planetary systems with low frequency observations

How do we sample the stellar weather and energetic particle events in these exoplanetary systems? In our solar system, CMEs often produce radio emission at low frequencies, known as Type II and III radio bursts. The emission is generated at the fundamental and first harmonic of the plasma frequency, while the flux density is related to the properties of the shock at the front of the CME which accelerates energetic particles [31]. As the radio plasma expands away from the Sun, the frequency rapidly declines such that these radio bursts are not visible from the ground when they are beyond a few solar radii.

No radio bursts have yet been detected from M dwarfs even though slowly rotating ones produce superflares that are 100 times larger than the Sun ([32]; see Figure 6). It is possible that the Alfvén speed near an M dwarf is high so that shocks only form at much greater distances from the star where the Alfvén speed drops below the CME velocity. In such circumstances, Type II radio bursts would only be visible below 10 MHz and undetectable from the ground. We have designed a lunar farside low radio frequency array (see below) to detect the equivalent of the brightest expected Type II and III bursts out to 10 pc at $\nu$<2 MHz. With this array, we will image the sky once per minute and monitor nearby stellar systems for large CMEs and energetic particle events.

### (b) Exoplanet magnetospheres

Radio emissions from planets with magnetic fields in the solar system are generated by electron cyclotron maser instabilities with electron gyro frequencies of $\sim 2.8 \times B$ MHz, where B is the magnetic field strength (G). Only Jupiter has a B-field strong enough for its radio emission to be detected from the ground. All the other planets with magnetic fields, including the Earth's Auroral Kilometric Radiation (AKR), are at $\nu \lesssim 2$ MHz. Detecting similar radio emission from exoplanets, below the Earth's ionospheric plasma frequency, is the only known way to probe for magnetospheres possibly needed for life to exist.

Detecting exoplanet auroral kilometric emission will require an array with a large collecting area at low frequencies. Recently, there has been a detection of radio emission from a large, free-floating 12.7±1 $M_{Jupiter}$ body [33] providing some encouragement for measuring B-fields associated with Earth-like planets. But, this will require a radio interferometer operating at ~MHz frequencies on the lunar farside.

Radio emission from nearby exoplanets depends upon the stellar wind conditions from the parent star. Extrapolations to potentially habitable extrasolar planets around M dwarfs predict radio powers that are orders of magnitude greater than the Earth [34]. For the Earth, enhanced solar wind conditions result in boosting the AKR by orders of magnitude and the same is expected for exoplanets. So, it will be important to image many nearby planetary systems simultaneously to detect when Type II radio bursts and AKR emission are enhanced.

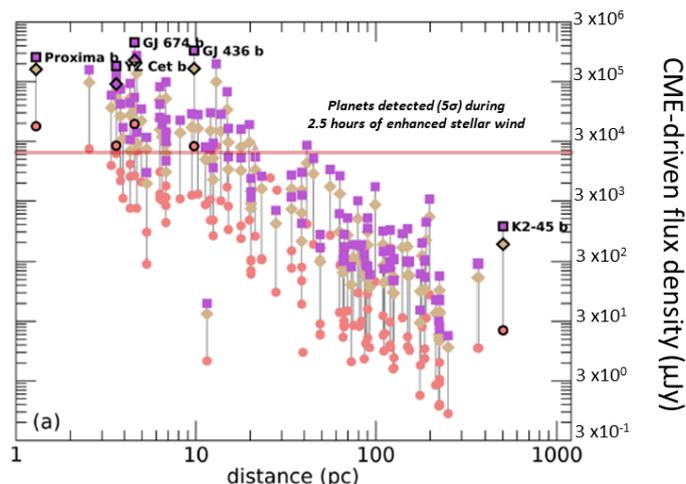

**Figure 7.** The predicted average flux density at 280 kHz of known exoplanets orbiting M dwarf stars assuming a magnetic field of 0.1 G. The color data points represent different models for the quiescent stellar wind. The horizontal line illustrates the 5$\sigma$ detection limit in a 2.5 hour integration during enhanced stellar wind conditions (e.g., CME). Adapted from [34]; courtesy of G. Hallinan, Caltech.

## (c) **FARSIDE:** Farside Array for Radio Science Investigations of the Dark ages and Exoplanets

We have developed a mission concept for a low radio frequency interferometer on the lunar farside to meet the science requirements outlined above. Here, we provide a high-level summary of the design and refer the reader to [35] for details on the array deployment, electronics, correlator, data analyses, etc. FARSIDE will image 10,000 deg² from the lunar farside every minute with 1400 channels spread over $\Delta v$=100 kHz – 40 MHz. In each sky image, there will be ~2000 planetary systems within 25 pc. FARSIDE will have sufficient resolution (a few degrees limited by scattering from the ISM and interplanetary medium) to identify which planetary system is undergoing a radio burst. Figure 7 illustrates how FARSIDE would detect radio emission from Earth and super-Earth planets around M dwarf stars in the solar neighborhood. This would provide the first measurements of planetary magnetic fields beyond the solar system, which would enhance the possibility of life, making these planets the objects of intense searches for biosignatures.

Ground-based arrays such as LOFAR [36], MWA [37], and OVRO-LWA [38] have blazed a trail in low frequency astrophysics that will be utilized by FARSIDE. The operational methodology, all-sky imagery, dynamic spectra, array calibration, and FPGA-based signal processing employed by these Earth-based arrays all feed-forward to FARSIDE (see [35]).

FARSIDE must be located on the farside of the Moon to avoid RFI and AKR from Earth, but also the wake cavity on the night side shields the radio antennas from system noise at <1 MHz due to electrons in the solar wind inducing currents [39]. With this environment, the design of the array is driven by the sensitivity and resolution which are listed in Table 1. 128 antenna nodes provide the required sensitivities in the low (0.1-2

MHz) band to meet the exoplanet observational goals. The effective collecting area of the array at 200 kHz using dipole antenna elements is an impressive 12.6 km². The spatial resolution is attained with a maximum baseline of 10 km deployed in a petal configuration as shown in Figure 8.

FARSIDE has a central base station emplaced on or near the lander along with two eMMRTGs (enhanced Multi-Mission Radioisotope Thermoelectric Generator) to power the array. The base also provides signal processing and telecommunications back to Earth via a relay satellite. The antenna nodes are deployed along eight independent spokes with a tether connecting the nodes in each spoke, distributing power and providing communications back to the base station. One or more rovers would carry antenna nodes out from the base station, unreeling the tether, before returning to base along a different path, then continuing to unreel the tether. Each node contains a short stacer antenna (5-m tip to tip) for the high frequencies and a long antenna (100-m tip to tip) built into the tether for low frequencies. The antenna node will be placed on the surface so the antenna axes are parallel throughout the array. Each antenna has a preamplifier, sending the two signals back to the base station via optical fibers as an analog signal.

**Table 1.** Baseline FARSIDE specifications.

| Quantity | Value |
|---|---|
| Antennas | 128 × 100 m length dipoles (100 kHz – 2 MHz), 128 × 5 m length dipoles (1-40 MHz) |
| Frequency Coverage | 100 kHz – 40 MHz (1400 × 28.5 kHz channels) |
| Field of View (FWHM) | > 10,000 deg² |
| Spatial Resolution | 10 degrees @ 200 kHz / 10 arcminutes @ 15 MHz |
| Antenna efficiency | $6.8 \times 10^{-6}$ @ 200 kHz / $9.5 \times 10^{-5}$ @ 15 MHz |
| System Temperature[a,b] | $1.0 \times 10^{6}$ K @ 200 kHz / $2.7 \times 10^{4}$ K @ 15 MHz |
| Effective Collecting Area[c] | ~ 12.6 km² @ 200 kHz / 2,240 m² @ 15 MHz |
| System Equivalent Flux Density (SEFD) | 230 Jy @ 200 kHz / $2.8 \times 10^{4}$ Jy @ 15 MHz |
| 1σ Sensitivity[b] (60 seconds; bandwidth = $\nu$/2) | 93 mJy @ 200 kHz[d] / 1.3 Jy (1.2 K) @ 15 MHz |
| 1σ Sensitivity[b] (1 hour; bandwidth = $\nu$/2) | 12 mJy @ 200 kHz[d] / 170 mJy (160 mK) @ 15 MHz |
| 1σ Sensitivity[b] (1000 hours; bandwidth = $\nu$/2) | 230 μJy[e] @ 200 kHz[d] / 3.8 mJy (5.2 mK) @ 15 MHz |

[a] *System temperature includes contribution from the sky and ground due to the absence of a ground screen.*
[b] *These values have been updated from the Astro 2020 report [40] due to increased fidelity in the front-end design.*
[c] *Effective area is impacted by loss of gain into the ground due to absence of a ground screen. Antenna efficiency not included.*
[d] *Sensitivity calculations at 200 kHz assume night time conditions.*
[e] *Deep confusion-free integrations are possible < 3 MHz due to the absence of extragalactic sources.*

The base station also houses the correlator. The integrated output of the correlator is passed to a radiation-tolerant, FPGA-based onboard computer and stored for relay to Earth. FARSIDE would collect full cross-correlation data every 60 seconds, in 1400 channels of 28.5 kHz width, for a total data rate of 65 GB per 24-hour period. Visibility data output would be transferred to Earth via the orbiting relay. A 10,000 deg² snapshot image would be produced for each 1400 channels in each band. Dynamic spectra at the location of each exoplanet system within 25 pc would be produced from these data to search for Type II/III radio bursts and planetary AKR.

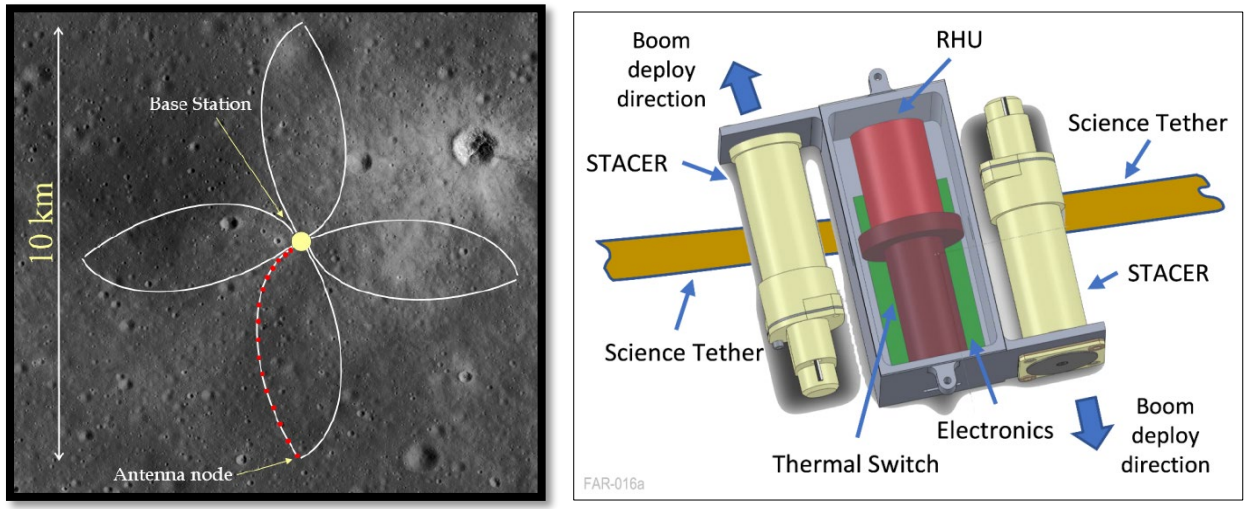

**Figure 8.** *Left:* FARSIDE will consist of 128 antenna nodes (red dots) deployed by a rover from a central base station, arranged in a petal configuration. *Right:* Mechanical design of the FARSIDE antenna node. The low frequency dipole is in the tether while the high frequency STACER dipole deploys orthogonally. RHU stands for radioisotope heater unit to permit survival and operations throughout the lunar night.

We baselined the transport to the lunar surface using the Blue Origin Blue Moon lander. The rover(s) rides on top of the lander so the entire assembly fits in a single launch vehicle fairing. The FARSIDE mission operation phases are shown schematically in Figure 9.

## (d) FARSIDE as a Dark Ages power spectrum pathfinder

FARSIDE would also be used as a pathfinder for observations of the Dark Ages 21-cm power spectrum. In many ways, spatial fluctuations in the 21 cm absorption during the cosmic Dark Ages provide the ultimate cosmological observable. The simplest way to quantify these fluctuations is with the power spectrum, which characterizes the amplitude of the variations as a function of spatial scale. During this time, the 21 cm line traces the cosmic density field with most modes in the linear or mildly non-linear regime, allowing a straightforward interpretation of the measurement in terms of the fundamental parameters of our Universe [41].

There are many challenges to detecting the Dark Ages power spectrum. Intrinsically, it is an extremely faint signature, and sky-noise dominates any measurement. The signal strength predicted by standard cosmological theory will require a collecting area of ~5 km² in the high band to be deployed above the Earth's ionosphere—far larger than the proposed FARSIDE array. Foreground emission complicates the picture even further. If residual foreground emission renders some modes unusable for cosmology, even more collecting area will be required to obtain a high significance 21-cm signal detection. FARSIDE is therefore an invaluable pathfinder for a larger 21 cm power spectrum instrument which can be constructed by adding more antenna nodes to the initial FARSIDE array. It can both measure the intrinsic spectrum of the low-frequency foregrounds without the effects of the Earth's ionosphere and can be used as a testbed for foreground removal techniques. Such measurements will be necessary for determining the array size ultimately necessary for a 21-cm Dark Ages power spectrum measurement.

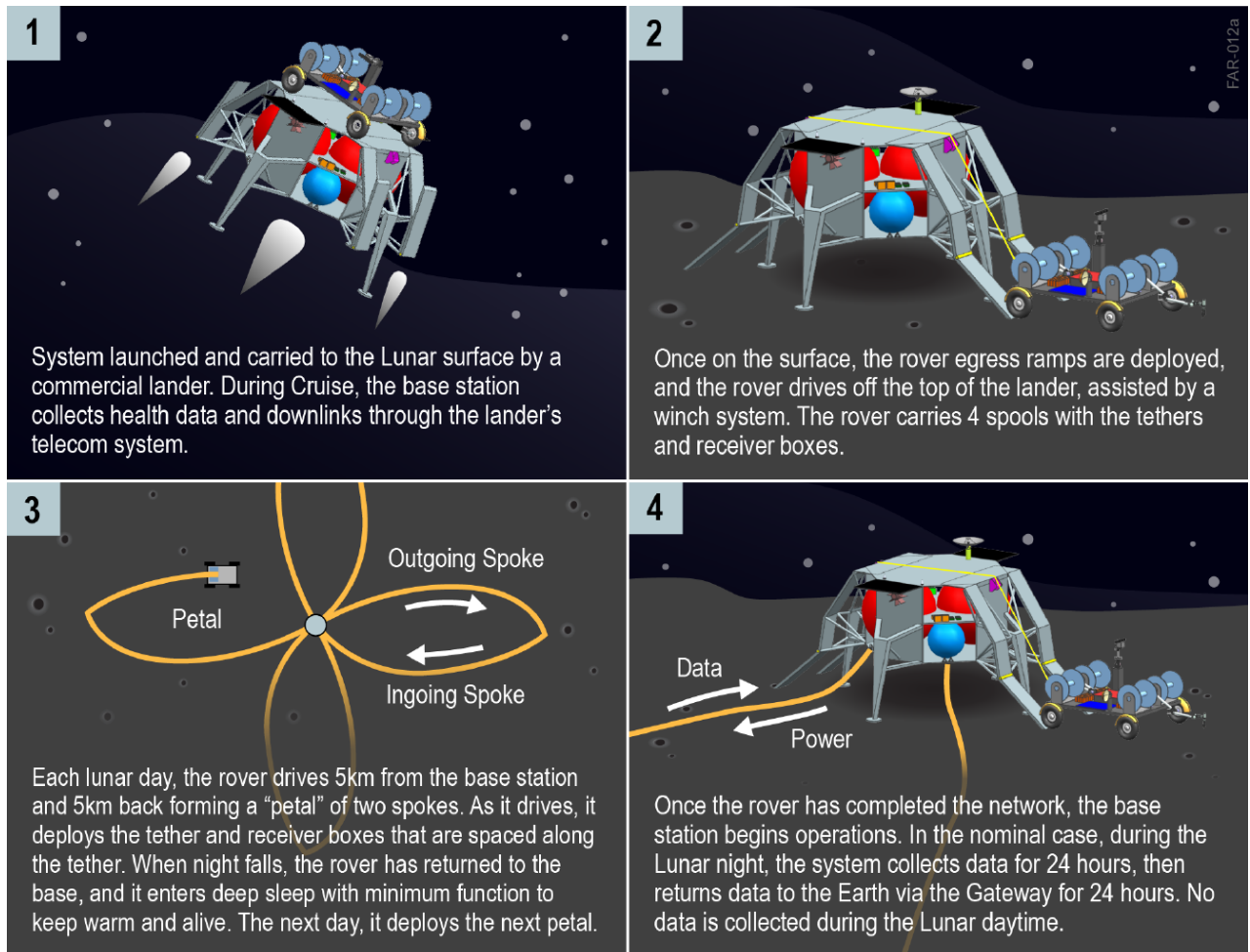

**Figure 9.** The FARSIDE mission has four phases: 1) Cruise; 2) Initial Surface operations, deploying the rover; 3) Spoke deployment—where the eight spokes are deployed as four petals; and 4) Science Operations.

# 4. Discussion and conclusions

NASA, along with all the major national space agencies, is committed to new science and resource explorations of the Moon in this decade. The NASA Commercial Lunar Payload Services (CLPS) program will deliver science payloads to the surface of the Moon beginning in 2021 [42]. Two of the CLPS landers will carry low radio frequency radio astronomy antennas. ROLSES, led by R. MacDowall at NASA GSFC, will measure the photoelectron sheath density of the lunar ionosphere near the surface and measure radio emissions from the Sun and outer planets. LuSEE, led by S. Bale at the University of California at Berkeley, will measure DC electric and magnetic fields also near the lunar surface. Both experiments will lay the groundwork for more challenging low frequency observations on the lunar farside.

DAPPER and FARSIDE are designed to take advantage of the emerging transportation and communication infrastructure associated with NASA's Artemis missions. DAPPER makes use of ride-share opportunities for a launch to the Moon and then descends to low lunar orbit using its spacecraft propulsion system. Once in orbit, DAPPER will make the first observations of the Dark Ages of the early Universe using the highly redshifted 21-cm signal, with data collected when the spacecraft is above the radio-quiet lunar farside and relayed to Earth via NASA communication assets. It will test the standard cosmological model in an unexplored epoch and probe for new physics, possibly involving novel interactions between baryons and dark matter.

FARSIDE is an array of 128 dipole antennas deployed from a commercial lander using one or more rovers. FARSIDE will have sufficient sensitivity below a few MHz to detect radio bursts associated with coronal mass ejections and other energetic space weather events from stars out to 25 pc. Such space plasma events from nearby stars would energize auroral kilometric radiation and, thus, provide a measure of the strength of magnetic fields associated with potentially habitable planets. Low frequency radio observations from the Moon are the only way to make such measurements of planetary B-fields, which may shield atmospheres from being stripped by stellar winds and preserve life.

This decade offers a unique opportunity for RF-quiet measurements from the lunar farside, as rideshare access and communications infrastructure become ubiquitous, and before further development of lunar assets compromises the radio-quiet character in the long-term. Thus, it is important for DAPPER and FARSIDE to launch in the next 5-10 years to seize the opportunity to (1) resolve fundamental questions about the validity of the ΛCDM model and the nature of the possible new physics in the Dark Ages, and (2) constrain space weather and planetary magnetic fields associated with exoplanets as a gateway to better understand the prospects of habitability in nearby star systems.

**Acknowledgments**

I am deeply grateful for all the work performed by the DAPPER and FARSIDE teams described within this paper. In particular, I wish to thank R. Bradley, S. Bale, D. Rapetti, K. Tauscher, B. Nhan, D. Bordenave, N. Bassett, J. Hibbard, J. Mirocha, S. Furlanetto, J. Pober, and J. Bowman from the DAPPER team. From FARSIDE, I am indebted to G. Hallinan, L. Teitelbaum, J. Lux, A. Romero-Wolf, T-C Chang, R. MacDowall, and the JPL Team-X for their creative efforts in designing the array. I also thank the leadership of the NASA Ames Research Center and Caltech/JPL for their backing of DAPPER and FARSIDE, respectively, and S. Squyres and the Blue Moon team at Blue Origin for their collaboration.

**Funding Statement**

This work is directly supported by the NASA Solar System Exploration Virtual Institute cooperative agreement 80ARC017M0006.

**Competing Interests**

I have no competing interests.